\documentclass[%
 utf8, 
 reprint,
 amsmath,
 amssymb,
 aps,
]{revtex4-2}

\usepackage[utf8]{inputenc}
\usepackage[T1]{fontenc}
\usepackage{graphicx}
\usepackage{dcolumn}
\usepackage{bm}
\usepackage{color}
\usepackage{cases}
\usepackage{textcomp}

\begin{document}

\title{Energy relaxation via quantum thermalization: A superconducting qubit coupled to an interacting many-body two-level system}


\author{Xue-Yi Guo}
\email{guoxy@baqis.ac.cn}
\affiliation{
 Beijing Academy of Quantum Information Sciences, Beijing 100193, China
}%

\date{\today}

\begin{abstract}

Two-level systems (TLS) are widely recognized as a dominant source of energy relaxation in superconducting qubits, limiting coherence times and device performance. 
While TLS-induced dissipation is often modeled as independent channels, interacting TLS can form finite many-body environments whose dynamical role in qubit relaxation remains insufficiently understood.
Building on the information-erasure-based framework for irreversible thermalization developed in our previous work, we present a microscopic model and numerical simulations of a superconducting qubit coupled to a finite interacting many-body TLS environment as a concrete physical realization of this mechanism. 
We show that effective irreversible energy relaxation can emerge from intrinsic many-body dynamics without invoking external baths or phenomenological dissipation.
The introduction of information erasure drives a transition from coherent oscillatory energy exchange to nonoscillatory exponential relaxation of the qubit. 
The relaxation times exhibit the scaling $T_1, T_2 \propto J_{q,\tau}^{-2}$ with respect to the qubit-TLS coupling strength, consistent with Fermi's golden rule and Lindblad-type descriptions. 
Importantly, both exponential decay and its scaling behavior arise directly from microscopic many-body dynamics.
These results provide quantitative insight into TLS-induced decoherence mechanisms and establish a microscopic link between information erasure, many-body thermalization, and irreversible relaxation in superconducting quantum systems.

\end{abstract}

\maketitle

\section{Introduction}

Superconducting qubits have demonstrated rapid progress toward scalable quantum information processing, including recent advances in small-scale quantum error correction \citep{acharya_quantum_2025, he_experimental_2025}. 
However, achieving long coherence times remains a central challenge, as superconducting qubits are inevitably coupled to complex material environments containing a variety of microscopic degrees of freedom. 
These include local defects originating from substrates, interfaces, and fabrication residues, as well as nonlocal excitations such as phonons and quasiparticles \citep{bilmes_resolving_2020, odeh_non-markovian_2025, barends_minimizing_2011, gustavsson_suppressing_2016, graaf_two-level_2020}. 
Interactions between qubits and these microscopic environments induce energy relaxation and decoherence, posing a major obstacle to scalable quantum computation. 
Understanding the microscopic origin of such decoherence mechanisms is therefore essential for improving qubit performance and enabling reliable quantum processors.

Local microscopic defects in superconducting devices are commonly modeled as two-level systems (TLS) \citep{phillips_tunneling_1972, phillips_two-level_1987, muller_towards_2019}. 
Extensive experiments on superconducting qubits and resonators have established several key properties of TLS environments. 
First, TLS are widely distributed in surface oxides, material interfaces, and fabrication residues\citep{gao_experimental_2008, wang_improving_2009, woods_determining_2019, crowley_disentangling_2023}. 
Second, TLS exhibit broad frequency distributions and can couple directly to superconducting qubits through electric-dipole interactions, enabling coherent energy exchange under strong coupling conditions \citep{lisenfeld_observation_2015, lisenfeld_electric_2019}. 
Third, interactions among TLS can induce frequency diffusion and spectral wandering, leading to temporal fluctuations of qubit relaxation times that are widely observed experimentally \citep{shalibo_lifetime_2010, meisner_probing_2018, carroll_dynamics_2022, berritta_real-time_2026}. 
These observations indicate that realistic TLS environments form interacting many-body systems rather than independent dissipative channels.

A common theoretical assumption in modeling TLS-induced qubit relaxation is that each TLS dissipates energy rapidly through phonon channels \citep{muller_relaxation_2009, cho_simulating_2023}. 
Within this framework, exponential qubit relaxation is typically described using Lindblad master equations or Fermi's golden rule, where dissipation is introduced phenomenologically. 
However, recent advances in phononic engineering have demonstrated that TLS lifetimes can be significantly prolonged, enabling regimes where TLS dissipation is strongly suppressed \citep{odeh_non-markovian_2025}. 
In such regimes, coherent qubit-TLS dynamics may dominate, and the microscopic origin of irreversible relaxation becomes less clear.

This raises a fundamental question: How can irreversible exponential relaxation emerge from coherent microscopic dynamics in a finite interacting TLS environment? 
Recent theoretical work has shown that in systems containing many lossless TLS, the quantum recurrence time increases exponentially with system size, leading to effective exponential decay on short timescales \citep{karimi_blueprint_2026}. 
However, the role of multiexcitation dynamics and many-body interactions in realistic TLS environments remains largely unexplored. 
In particular, it remains unclear how internal thermalization within a finite interacting TLS influences qubit energy relaxation and decoherence.

In our previous works \citep{guo_thermalization_2025, guo_thermalization2_2025}, we proposed a microscopic dynamical mechanism underlying the irreversibility of the second law of thermodynamics, namely, information erasure. 
The central idea is that, during thermalization, while the unitary dynamics of an isolated system continuously preserves information and tends to recover the initial state, information erasure breaks information conservation and drives the system toward equilibrium.
Therefore, entropy does not increase spontaneously with time; rather, entropy production relies on information erasure. This provides a different perspective from the conventional interpretation of the second law, in which entropy increase is regarded as an intrinsic tendency of isolated systems under time evolution.
In quantum mechanics, time evolution is always linear and unitary, but it is by no means trivial. Its complexity is analogous to that of the three-body problem in classical mechanics. For a single particle confined in a box, the wave function undergoes coherent spreading and revival. For two particles, interactions generate entanglement between them; however, in a finite-size system, subsequent collisions can disentangle the particles, allowing the system to return to its initial state. For three particles, the collision of the third particle continuously perturbs the entanglement and disentanglement dynamics between the other two particles, giving rise to chaotic quantum dynamics analogous to the classical three-body problem.
Therefore, information erasure can naturally emerge in complex quantum systems and provide the microscopic origin of thermalization. 
In summary, in Refs.~\citep{guo_thermalization_2025, guo_thermalization2_2025}, we demonstrated, using a three-component fermionic Hubbard system, that entanglement and entropy increase, while subsystem mutual information decreases, under an information-erasure sequence. These results show that the second law of thermodynamics can emerge solely from quantum mechanics without invoking additional assumptions.

In this work, we demonstrate that irreversible qubit energy relaxation can emerge naturally from intrinsic information-erasure processes in a finite interacting many-body TLS environment undergoing internal thermalization. 
By constructing a microscopic dynamical model of a superconducting qubit coupled to interacting TLS, we perform numerical simulations that capture the transition from coherent qubit-TLS energy exchange to exponential decay as the TLS subsystem thermalizes. 
We further show that the resulting relaxation time depends sensitively on both the qubit-TLS coupling strength and intrinsic TLS properties, such as interaction strength, frequency fluctuations, and thermal excitation. 
These results establish a microscopic mechanism linking intrinsic information erasure and many-body thermalization to irreversible relaxation and provide insight into experimentally observed fluctuations of qubit relaxation times.

Beyond superconducting qubits, the mechanism identified here may provide a general framework for understanding irreversible relaxation processes in other finite interacting quantum systems, where irreversibility emerges from intrinsic many-body dynamics rather than external baths. 
The remainder of this paper is organized as follows. 
In Sec.~II, we introduce the model and parameter settings. 
In Sec.~III, we present numerical results and analyze the dependence of relaxation dynamics on system parameters. 
Finally, we summarize the main conclusions and discuss future directions.

\section{Model}

\begin{figure}[htbp]  
	\centering  
	\includegraphics[scale=1.0]{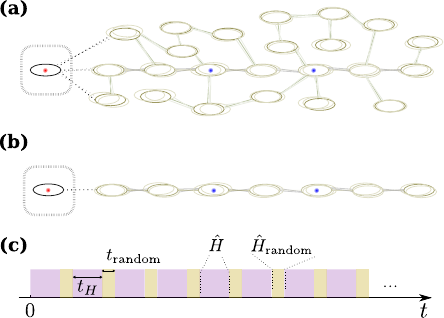}    
\caption{\textbf{Many-body TLS model and the full forward-time information-erasure (FFTIE) sequence.}
(a) Schematic illustration of the coupled superconducting qubit-TLS network. 
The black circle (dashed box) represents the qubit, and the gold circles represent TLS. 
Solid lines denote near-resonant TLS-TLS couplings, while the dashed line denotes weak near-resonant qubit-TLS coupling. 
The ghosting effect illustrates TLS with distinct frequencies, and spheres of different colors represent their excited states. 
(b) One-dimensional TLS chain used in numerical simulations, capturing the essential many-body interactions responsible for internal thermalization. 
(c) FFTIE sequence employed in the simulations, which drives internal thermalization of the TLS subsystem and enables the emergence of irreversible qubit relaxation.
}\label{fig1}
\end{figure}

We consider a superconducting qubit coupled to a finite interacting many-body two-level system, as illustrated in Fig.~\ref{fig1}(a). 
The qubit represents an artificial atom in a superconducting circuit, while the TLS correspond to microscopic defects commonly found in amorphous materials, interfaces, and substrates. 
Interactions between TLS arise from electric dipole-dipole coupling or phonon-mediated processes, both of which have been observed experimentally \citep{lisenfeld_observation_2015, lisenfeld_decoherence_2016}. 
Such interactions naturally lead to the formation of an interacting many-body TLS environment.

Following the information-erasure framework introduced in Refs.~\cite{guo_thermalization_2025, guo_thermalization2_2025}, we construct an interacting many-body TLS by classifying TLS into two types, denoted as $\tau$ and $\upsilon$, corresponding to defects with different characteristic frequencies. 
Nearest-neighbor TLS with similar frequencies are coupled through resonant hopping interactions, while distinct-frequency TLS interact through effective density-density (ZZ-type) couplings.

We model the TLS as a one-dimensional chain, as shown in Fig.~\ref{fig1}(b). 
Although realistic devices contain disordered three-dimensional TLS networks, a short one-dimensional chain provides a minimal building block that already captures many-body interaction effects while remaining computationally tractable. 
In the main text, we focus on a chain length of $L_{\mathrm{TLS}} = 7$. Numerical results for $L_{\mathrm{TLS}} = 3$, $5$, and $9$ are presented in the Appendix. Although $L_{\mathrm{TLS}}$ is relatively modest, the resulting many-body dynamics is already sufficiently complex to capture the onset of thermalization.
Under these assumptions, the TLS subsystem can be described by a two-component Hubbard-type Hamiltonian
\begin{align}\label{tlsH0}
	\hat{H}_{\mathrm{TLS}} &= \sum_{\langle i,j \rangle} 
	\Big[
	J_\tau 
	\left( 
	\hat{c}_{i,\tau}^\dagger \hat{c}_{j,\tau} 
	+ 
	\hat{c}_{i,\tau} \hat{c}_{j,\tau}^\dagger 
	\right)
	\notag \\
	&\quad +
	J_\upsilon 
	\left(
	\hat{c}_{i,\upsilon}^\dagger \hat{c}_{j,\upsilon}
	+
	\hat{c}_{i,\upsilon} \hat{c}_{j,\upsilon}^\dagger
	\right)
	\Big]
	\notag \\
	&\quad +
	\sum_i 
	\left[
	U_{i,\tau}\hat{n}_{i,\tau}
	+
	U_{i,\upsilon}\hat{n}_{i,\upsilon}
	\right]
	\notag \\
	&\quad +
	\sum_i 
	U_{\tau,\upsilon}
	\hat{n}_{i,\tau}
	\hat{n}_{i,\upsilon}.
\end{align}

The superconducting qubit is assumed to be resonantly coupled to the leftmost $\tau$-type TLS with coupling strength $J_{q,\tau}$. 
The total Hamiltonian of the combined system is therefore
\begin{equation}\label{H0}
	\hat{H}
	=
	U_q \hat{n}_q
	+
	\hat{H}_{\mathrm{TLS}}
	+
	J_{q,\tau}
	\left(
	\hat{c}_q^\dagger \hat{c}_{0,\tau}
	+
	\hat{c}_q \hat{c}_{0,\tau}^\dagger
	\right).
\end{equation}

A key ingredient of the present model is the inclusion of stochastic TLS frequency fluctuations. 
Physically, such fluctuations arise from uncontrolled microscopic environments and interactions with phonon modes, and are widely observed experimentally. 
To incorporate this effect, we introduce intermittent evolution segments governed by a random Hamiltonian
\begin{equation}
\hat{H}_{\mathrm{random}}
=
\sum_i
U_{i,\upsilon,\mathrm{random}}
\hat{n}_{i,\upsilon},
\end{equation}
where each $U_{i,\upsilon,\mathrm{random}}$ is randomly drawn for each realization.
We further emphasize that no experimental system is perfectly isolated. In practice, heat and particle exchange with the environment can be strongly suppressed, while the environment can still induce adiabatic phase perturbations that effectively serve as information-erasure processes.

The total time evolution consists of alternating segments governed by $\hat{H}$ and $\hat{H}_{\mathrm{random}}$, as illustrated in Fig.~\ref{fig1}(c). 
Specifically, we define
\begin{align}
\hat{U}(t_i)
&=
e^{-i\hat{H} t_H}, 
\\
\hat{O}_i
&=
e^{-i\hat{H}_{\mathrm{random}} t_{\mathrm{random}}},
\end{align}
with independent realizations of $\hat{H}_{\mathrm{random}}$ for each $\hat{O}_i$.

This forward-time stochastic sequence is closely related to the information-erasure-based thermalization protocols introduced in Refs.~\citep{guo_thermalization_2025, guo_thermalization2_2025}. 
Specifically, the sequence shown in Fig.~\ref{fig1}(c) can be obtained by setting the time arguments in the information-erasure sequence of Refs.~\citep{guo_thermalization_2025, guo_thermalization2_2025} to the arithmetic progression
$[\cdots \hat{O}_i(i \times t_H)\cdots \hat{O}_2(2\times t_H)\hat{O}_1(t_H)]$, where $\hat{O}_i(t) = e^{i\hat{H} t} \hat{O}_i e^{-i\hat{H} t}$.
In contrast to the original information-erasure sequence, which may include reverse-time operations, the present model employs purely forward-time evolution to directly investigate irreversible relaxation processes. We therefore refer to this implementation as the FFTIE sequence.

Physically, the FFTIE sequence models stochastic frequency drift of TLS within the interacting many-body TLS environment. 
In our previous works \citep{guo_thermalization_2025, guo_thermalization2_2025}, we showed that in interacting many-body systems, irreversible dynamics can naturally emerge from intrinsic information-erasure processes, without requiring external baths or phenomenological dissipation. 
In the present context, the stochastic fluctuation of TLS frequencies provides a concrete microscopic realization of intrinsic information erasure, in which the system wave function progressively evolves into a superposition involving an increasing number of accessible microstates due to multicomponent interactions.
The effective information erasure within the many-body TLS leads to the progressive breakdown of coherent qubit-TLS energy exchange and gives rise to irreversible exponential relaxation of the qubit energy.

\section{Numerical Simulation Results}

In the total Hamiltonian $\hat{H}$, we set $U_q/2\pi = 0~\mathrm{MHz}$ and vary $J_{q,\tau}/2\pi$ from $0.002$ to $0.1~\mathrm{MHz}$. For the TLS chain, we fix $J_{\tau}/2\pi = J_{\upsilon}/2\pi = 1~\mathrm{MHz}$, $U_{i,\tau}/2\pi = U_{i,\upsilon}/2\pi = 0~\mathrm{MHz}$, and $U_{\tau,\upsilon}/2\pi = -0.2~\mathrm{MHz}$.
In $\hat{H}_{\mathrm{random}}$, each $U_{i,\upsilon,\mathrm{random}}/2\pi$ is drawn independently from a uniform distribution on $[0, 3]~\mathrm{MHz}$.

\begin{figure}[htbp]
	\centering
	\includegraphics[width=\columnwidth]{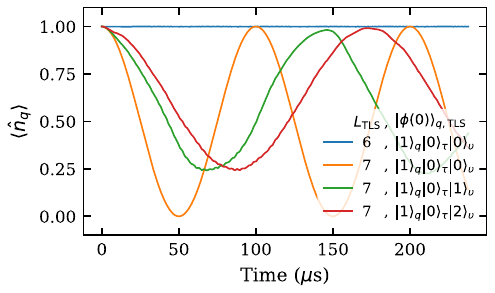}
	\caption{
		\textbf{Coherent energy exchange between the qubit and the TLS chain.}
Time evolution of the qubit occupation $\langle \hat{n}_q \rangle$ for many-body TLS chain lengths $L_{\mathrm{TLS}} = 6$ and $7$, starting from different initial states $|\phi(0)\rangle_{q,\mathrm{TLS}}$.
	}
	\label{fig2}
\end{figure}

When the internal coupling within the TLS chain is strong compared with the qubit-TLS coupling $J_{q,\tau}$, the finite chain length leads to a discrete many-body energy spectrum. 
As a result, the resonance condition between the qubit and the TLS becomes highly sensitive to the chain length. 
For example, with $L_{\mathrm{TLS}}=6$ and $J_{q,\tau}/2\pi=0.01~\mathrm{MHz}$, the state $|1\rangle_q |0\rangle_\tau|0\rangle_\upsilon$ is approximately an eigenstate of $\hat{H}$. 
Consequently, the qubit occupation $\langle \hat{n}_q \rangle$ remains nearly constant, as shown by the blue curve in Fig.~\ref{fig2}. 

In contrast, for $L_{\mathrm{TLS}}=7$, the same initial state becomes a superposition of coupled-system eigenstates, leading to complete coherent energy exchange between the qubit and the TLS chain, as illustrated by the orange curve in Fig.~\ref{fig2}. 
This coherent oscillation reflects the discrete nature of the finite TLS spectrum and serves as the baseline dynamical behavior prior to thermalization.

If the internal TLS coupling becomes comparable to $J_{q,\tau}$, the qubit excitation propagates along the chain in a manner analogous to a single-photon quantum walk. 
In this regime, the oscillation amplitude depends only weakly on the chain length. 
To investigate how energy is transferred into the TLS chain and subsequently redistributed between the qubit and the TLS chain over a broad range of qubit-TLS coupling strengths, including the weak-coupling regime, we adopt an odd TLS chain length of $L_{\mathrm{TLS}}=7$ throughout the main text. 
This choice ensures resonant energy transfer even for small values of $J_{q,\tau}$. Numerical results for $L_{\mathrm{TLS}}=3$, $5$, and $9$ are presented in Fig.~\ref{figA1} of the Appendix .

The number of TLS excitations also affects the coherent energy exchange. For a single $\upsilon$-type excitation, i.e., the initial state $|0000000\rangle_\tau|1000000\rangle_\upsilon$ (denoted succinctly as $|0\rangle_\tau|1\rangle_\upsilon$), the qubit exhibits the coherent oscillation shown by the green curve in Fig.~\ref{fig2}, with an increased period and decreased amplitude. When the $\upsilon$-type excitation number is increased to 2, the period further increases while the amplitude remains almost unchanged.
Since TLS excitation probabilities are low under typical low-temperature experimental conditions, we focus primarily on few-excitation regimes.

\begin{figure*}[htbp]
    \centering
    \includegraphics[width=\linewidth]{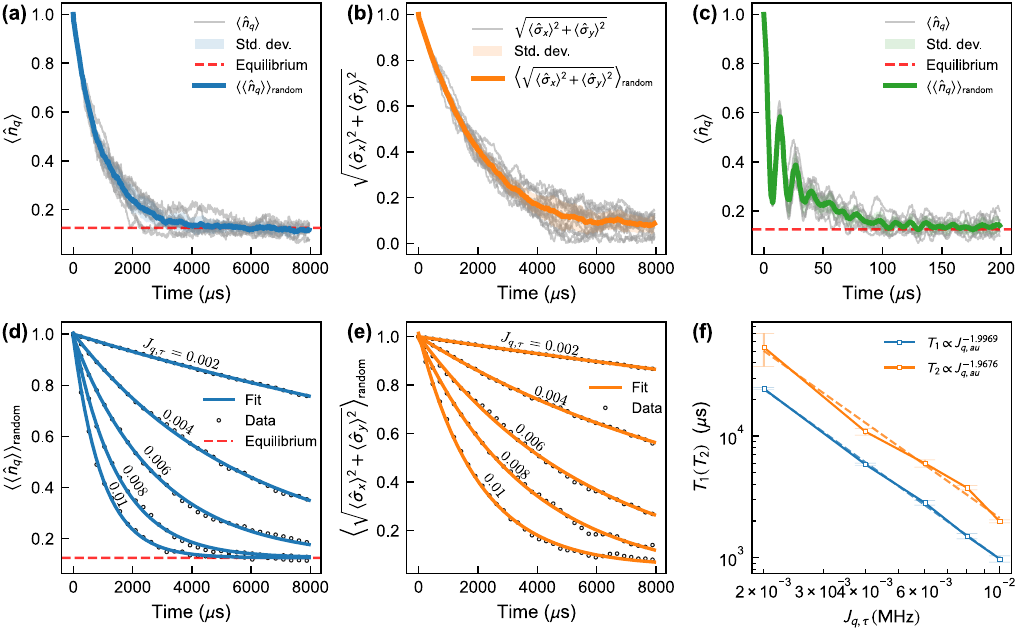}
    \caption{
\textbf{Exponential qubit relaxation and dephasing induced by many-body thermalization.}
(a), (b) Time evolution of the qubit occupation $\langle \hat{n}_q \rangle$ and coherence $\sqrt{\langle \hat{\sigma}_x \rangle^2 + \langle \hat{\sigma}_y \rangle^2}$ for $J_{q,\tau}=0.01$, with initial states $|1\rangle$ and $(|0\rangle+|1\rangle)/\sqrt{2}$, respectively. Solid blue (orange) lines represent averages over ten trajectories, shaded regions indicate standard deviations, and light-gray lines show individual trajectories.
(c) Time evolution of $\langle \hat{n}_q \rangle$ for $J_{q,\tau}=0.1$ (initial state $|1\rangle$). The solid green line represents the average over ten trajectories, the shaded region indicates the standard deviation, and the light-gray lines show individual trajectories. 
(d), (e) Time evolution of  $\langle \hat{n}_q \rangle$ and $\sqrt{\langle \hat{\sigma}_x \rangle^2 + \langle \hat{\sigma}_y \rangle^2}$ for $J_{q,\tau}=0.01, 0.008, 0.006, 0.004, 0.002$, with initial states $|1\rangle$ and $(|0\rangle+|1\rangle)/\sqrt{2}$, respectively. Solid lines are exponential fits, and open circles represent the averaged raw data (each average containing ten trajectories).
(f) Extracted $T_1$ and $T_2$ values as functions of $J_{q,\tau}$. The dashed lines are power-law fits to the data, with error bars derived from the fitting uncertainties of $T_1$ and $T_2$ in panels (d) and (e).
}\label{fig3}
\end{figure*}

Such coherent energy exchange between the qubit and TLS in superconducting quantum chips demands high error-correction rates (continuous quantum error correction) \citep{oreshkov_continuous_2007}, which is challenging to achieve experimentally. In contrast, we demonstrate that many-body TLS thermalization suppresses this coherent qubit-TLS exchange, transforming it into exponential energy relaxation---a feature beneficial for practical quantum error correction.
As shown in Fig.~\ref{fig3}, starting from $|\phi(0)\rangle_{q,\mathrm{TLS}} = |1\rangle_q |0\rangle_\tau |2\rangle_\upsilon$, the combined system evolves unitarily under the FFTIE sequence and remains in the pure state $|\phi(t)\rangle_{q,\mathrm{TLS}}$. With $J_{q,\tau}/2\pi=0.01~\mathrm{MHz}$, $t_H=2/2\pi~\mu\mathrm{s}$, and $t_{\mathrm{random}}=0.5/2\pi~\mu\mathrm{s}$, the qubit occupation $\langle \hat{n}_q \rangle$ exhibits exponential decay across individual trajectories [Fig.~\ref{fig3}(a), light-gray lines], each displaying consistent decay profiles---signifying irreversible thermalization. During this process, the TLS chain also thermalizes and converges to the same equilibrium state as the qubit (Fig.~\ref{figA3}).

The solid blue line in Fig.~\ref{fig3}(a) represents the average $\langle\langle \hat{n}_q \rangle\rangle_{\mathrm{random}}$ over ten individual evolutions, where $\langle \cdot \rangle_{\mathrm{random}}$ denotes averaging over different realizations of the random Hamiltonian sequence. 
Each individual trajectory already exhibits clear exponential relaxation, with small stochastic fluctuations arising from the random parameters in $\hat{H}_{\mathrm{random}}$. 
The averaged decay curve becomes significantly smoother and gradually saturates to the microcanonical equilibrium value of $1/8$. 
This equilibrium value is given by the ratio of the total number of excitations to the total number of lattice sites. 
Since the main text considers only the case of a single initial qubit excitation, the total number of excitations is 1, while the total number of sites, including the qubit and the TLS chain, is 8. Therefore, the equilibrium qubit occupation is $1/8$.
Numerical results for initial states containing $\tau$-type TLS excitations are presented in Fig.~\ref{figA2} of the Appendix. 
The light blue shaded region in Fig.~\ref{fig3}(a) indicates one standard deviation above and below the average.

It is worth noting that in experiments, obtaining the probability of the qubit being in $|1\rangle_q$ at time $t$ requires repeated state preparation, evolution, and measurement. 
Such procedures inherently correspond to averaging over multiple relaxation realizations. 
Consequently, the experimentally measured $T_1$ relaxation curve naturally corresponds to such averaged dynamics, making the averaged result in Fig.~\ref{fig3}(a) directly comparable to experimental observations.

Under the same FFTIE sequence, the relaxation dynamics of $\sqrt{\langle \hat{\sigma}_x \rangle^2 + \langle \hat{\sigma}_y \rangle^2}$ are shown in Fig.~\ref{fig3}(b) for the qubit initialized in $(|0\rangle+|1\rangle)/\sqrt{2}$. 
The light-gray lines (ten trajectories) each exhibit a monotonic decay consistent with irreversible thermalization dynamics. 
The solid orange line represents the average $\langle \sqrt{\langle \hat{\sigma}_x \rangle^2 + 
\langle \hat{\sigma}_y \rangle^2}\rangle_{\mathrm{random}}$ over the ten trajectories, displaying a pronounced exponential decay trend.

Figures~\ref{fig3}(d) and ~\ref{fig3}(e) present the numerical results of $\langle\langle \hat{n}_q \rangle\rangle_{\mathrm{random}}$ and $\langle \sqrt{\langle \hat{\sigma}_x \rangle^2 + \langle \hat{\sigma}_y \rangle^2}\rangle_{\mathrm{random}}$ 
(open circles) for $J_{q,\tau}/2\pi = 0.01$, $0.008$, $0.006$, $0.004$, and $0.002~\mathrm{MHz}$, with the corresponding exponential fits 
shown as solid lines.

In Fig.~\ref{fig3}(f), the extracted $T_1$ values (blue open squares) are $975.88 (\pm 9.04)$, $1501.01 (\pm 11.64)$, $2827.45 (\pm 20.26)$, $5846.01 (\pm 23.10)$, and $24536.80 (\pm 78.49)\ \mu\mathrm{s}$.  The smallest of these values is comparable to experimentally reported coherence times \citep{bland_millisecond_2025},  providing a useful reference for realistic device regimes. The extracted $T_2$ values (orange open squares) are $1987.0 (\pm 9.1)$, $3758.5 (\pm 24.7)$, $5966.2 (\pm 64.8)$, $10880.5 (\pm 112.1)$, and $53755.6 (\pm 2600.3)\ \mu\mathrm{s}$.
Error bars represent fitting uncertainties, and the dashed lines are power-law fits to the data.

The fitted $T_1$ values scale as $J_{q,\tau}^{-1.9969}$, while $T_2$ scales as $J_{q,\tau}^{-1.9676}$. 
The nearly quadratic inverse scaling of both $T_1$ and $T_2$ with $J_{q,\tau}$ is consistent with the second-order dependence predicted by Fermi's golden rule for weak system-environment coupling. 
Importantly, in the present model this scaling emerges directly from unitary many-body dynamics without invoking an explicit Markovian bath or phenomenological Lindblad description.

The extracted $T_2$ values are approximately $2.14$ times $T_1$, consistent with relaxation-limited dephasing in the absence of additional pure-dephasing channels. 
In our model, the qubit is only weakly coupled to the TLS chain without additional noise sources; consequently, its dephasing originates solely from energy relaxation.

When $J_{q,\tau}$ is further increased, the qubit energy relaxation gradually transitions from exponential decay to oscillatory decay. 
Figure~\ref{fig3}(c) shows the numerical results for $J_{q,\tau}/2\pi=0.1~\mathrm{MHz}$. 
Both individual trajectories and the average exhibit clear oscillatory relaxation behavior, consistent with 
experimentally observed oscillatory relaxation in the strong qubit-TLS coupling regime \citep{lisenfeld_observation_2015}. 
The fitted $T_1$ value is $13.0 (\pm 0.2)\ \mu\mathrm{s}$, which is close to the extrapolated value of $9.8\ \mu\mathrm{s}$ obtained from Fig.~\ref{fig3}(f).

These results demonstrate that exponential qubit relaxation and golden-rule-like scaling can arise as emergent phenomena 
from intrinsic many-body thermalization dynamics, even in a finite interacting quantum environment.

\begin{figure}[htbp]
	\centering
	\includegraphics[width=\columnwidth]{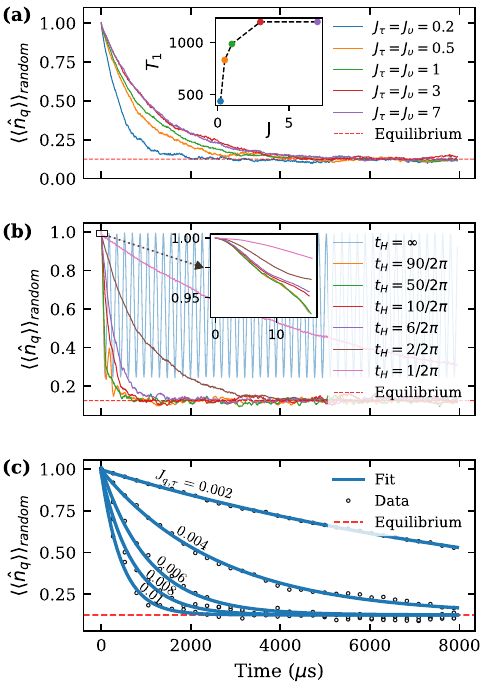}
\caption{
\textbf{Dependence of qubit relaxation on many-body TLS environment parameters.}
(a) Extracted $T_1$ averaged over five trajectories for varying internal TLS couplings $J_{\tau}$ and $J_{\upsilon}$, with other parameters identical to those in Fig.~\ref{fig3}(a).
(b) Extracted $T_1$ for different FFTIE interval times $t_H$, controlling the rate of effective TLS frequency fluctuations.
(c) Extracted $T_1$ scaling with $J_{q,\tau}$ for reduced TLS excitation number (initial state $|1\rangle_q|0\rangle_{\tau}|1\rangle_{\upsilon}$ with a single $\upsilon$ excitation). Dashed lines represent averaged data, and solid lines denote exponential fits.
}\label{fig4}
\end{figure}

The scaling behavior of $T_1$ with $J_{q,\tau}$ demonstrated in Fig.~\ref{fig3} indicates that, under fixed environmental conditions, the qubit relaxation time is primarily controlled by the qubit-TLS coupling strength. 
In realistic devices, however, TLS environments exhibit substantial variability. 
Parameters such as internal couplings can differ significantly across fabricated chips due to material disorder and fabrication variability, while frequency fluctuation rates and thermal occupation numbers depend on the experimental environment, particularly environmental temperature and its fluctuations \citep{shalibo_lifetime_2010}.
From the perspective of open quantum system theory, such variations effectively modify environmental correlation properties and thus alter the qubit relaxation rate. 
To explore these effects, we now investigate how different many-body TLS parameters influence qubit relaxation dynamics.

Figure~\ref{fig4}(a) shows the dependence of the qubit relaxation time on the internal TLS coupling strengths $J_{\tau}/2\pi$ and $J_{\upsilon}/2\pi$, which are set to $0.2$, $0.5$, $1$, $3$, and $7~\mathrm{MHz}$, respectively, while all other parameters remain identical to those in Fig.~\ref{fig3}(a). 
For sufficiently strong internal coupling ($J_{\tau}/2\pi, J_{\upsilon}/2\pi > 3~\mathrm{MHz}$), the extracted $T_1$ values saturate and become nearly constant. 
As $J_{\tau}$ and $J_{\upsilon}$ decrease, $T_1$ is progressively reduced and becomes increasingly sensitive to parameter changes, particularly in the regime below $J_{\tau}/2\pi, J_{\upsilon}/2\pi \approx 1~\mathrm{MHz}$.

This behavior can be interpreted from the structure of the  eigenstates of the interacting TLS subsystem. 
The number of many-body states that can participate in coherent energy exchange with the qubit plays a central role in determining the relaxation rate.
When the internal TLS coupling is relatively weak, the many-body spectrum remains narrow, and multiple eigenstates lie within the effective resonance window set by the qubit-TLS coupling strength. 
As a result, several coherent exchange pathways are available, allowing excitation energy to be distributed among multiple TLS modes and leading to faster relaxation.
As the internal TLS coupling increases, the many-body spectrum broadens, reducing the number of eigenstates that remain resonant with the qubit. 
Consequently, the number of effective exchange channels decreases, suppressing energy transfer and resulting in longer relaxation times.
Beyond a certain coupling strength, the number of resonant many-body states approaches a constant value, typically dominated by a single near-resonant mode. 
This leads to saturation of the relaxation time, consistent with the behavior observed in Fig.~4(a).

For $t_H = \infty$, the system exhibits persistent coherent energy exchange between the qubit and the TLS chain, 
with an oscillation period of approximately $1093/2\pi~\mu\mathrm{s}$. 
This behavior corresponds to the coherent exchange dynamics shown previously in Fig.~\ref{fig2} and serves as a baseline 
for comparison with the FFTIE-driven relaxation processes.
Notably, even at the minima of the oscillations, the qubit population remains significantly above the thermal equilibrium value. 
This indicates that coherent energy exchange alone, while facilitating energy redistribution, is insufficient to induce thermalization, highlighting the essential role of the FFTIE sequence in enabling irreversible relaxation dynamics.

When the unitary evolution segment $e^{-i\hat{H}_{\mathrm{random}} t_{\mathrm{random}}}$ is introduced at relatively long intervals, for example, $t_H = 90/2\pi~\mu\mathrm{s}$, coherent oscillations remain visible for several cycles before gradually decaying toward equilibrium. 
At $t_H = 50/2\pi~\mu\mathrm{s}$, coherent oscillations are almost completely suppressed. 
As $t_H$ decreases further, the relaxation curves deviate from the coherent oscillatory behavior from the earliest times 
[see inset in Fig.~\ref{fig4}(b)] and exhibit clear exponential relaxation at long times. 
These results demonstrate that increasing the FFTIE rate effectively suppresses coherent qubit-TLS oscillations and 
promotes irreversible relaxation dynamics, thereby stabilizing qubit relaxation behavior. 
Such suppression of coherent backflow is advantageous for maintaining predictable relaxation processes in quantum information applications.

Figure~\ref{fig4}(c) explores the effect of TLS excitation number. 
Here, the parameters are identical to those in Fig.~\ref{fig3}(d), except that the number of $\upsilon$-type TLS excitations is reduced from $2$ to $1$. 
The resulting relaxation curves for different $J_{q,\tau}$ yield fitted $T_1$ values of $459.5 (\pm 18.2)$, $714.6 (\pm 15.1)$, $1067.9 (\pm 26.9)$, $2637.6 (\pm 31.6)$, and $10402.6 (\pm 41.0)\ \mu\mathrm{s}$.
The extracted scaling behavior $T_1 \propto J_{q,\tau}^{-1.944}$ remains consistent with the results in Fig.~\ref{fig3}(d), indicating that the inverse quadratic scaling with qubit-TLS coupling strength is robust against changes 
in excitation number.

However, the absolute $T_1$ values are reduced by nearly half compared to those in Fig.~\ref{fig3}(d). 
Since TLS excitation probabilities are temperature dependent in experimental systems, this result suggests that changes 
in the initial excitation distribution of the TLS can significantly influence the effective relaxation rate. 
Interestingly, under certain parameter regimes, fewer excited TLS (i.e., lower effective temperatures) can lead to shorter qubit lifetimes, highlighting the nontrivial role of many-body excitation structure in determining relaxation behavior.

\section{Summary and outlook}

In this work, we have demonstrated that irreversible energy relaxation of a superconducting qubit can arise dynamically 
from the intrinsic evolution of a finite interacting many-body TLS environment. 
Within this microscopic model, internal thermalization of the TLS subsystem suppresses coherent qubit-TLS energy exchange 
and drives a crossover from coherent oscillations to exponential relaxation. 
This result provides a concrete dynamical mechanism linking many-body thermalization and information-erasure processes 
to effectively irreversible relaxation in finite quantum environments.

Our simulations further reveal how the relaxation dynamics depend on key microscopic parameters. 
Stronger internal TLS coupling, higher TLS fluctuation rates (FFTIE), and increased TLS excitation collectively suppress 
coherent oscillations and promote irreversible relaxation. 
In addition, the qubit relaxation time follows a scaling behavior $T_1 \propto J_{q,\tau}^{-2}$, consistent with 
Fermi's golden rule and Lindblad-type descriptions. 
These results provide a microscopic interpretation of experimentally observed fluctuations in qubit relaxation 
times \citep{klimov_fluctuations_2018, bland_millisecond_2025}, which remain a major challenge for quantum error correction \citep{etxezarreta_martinez_time-varying_2021, iolius_performance_2022, etxezarreta_martinez_multiqubit_2023}.

TLS on superconducting quantum chips can contribute to dissipation through multiple channels, including phonon-mediated relaxation and intrinsic many-body TLS dynamics. 
While phonon-assisted TLS dissipation has been widely studied and is often considered an important dissipation channel, 
recent work indicates that suppressing TLS-phonon coupling can significantly extend TLS lifetimes \citep{odeh_non-markovian_2025}. 
In such regimes, dissipation arising from interacting many-body TLS becomes increasingly relevant, making the mechanism identified here particularly significant.

The information-erasure framework developed in Refs.~\citep{guo_thermalization_2025, guo_thermalization2_2025} provides a general theoretical perspective for analyzing irreversible thermalization and quantum chaos in finite systems. 
The present work demonstrates how this framework can be applied to realistic qubit-TLS environments and used to interpret experimentally relevant relaxation phenomena.

Looking forward, advanced TLS control techniques \citep{lisenfeld_observation_2015, meisner_probing_2018, lisenfeld_electric_2019, bilmes_resolving_2020, hegedus_situ_2025, dane_robust_2026, berritta_real-time_2026} provide promising experimental routes to test the predictions presented here. 
For example, tuning the number of excited TLS \citep{odeh_non-markovian_2025} or modifying internal TLS coupling strengths could enable controlled observation of the predicted crossover from coherent oscillations to exponential relaxation. 
From a theoretical perspective, the one-dimensional TLS chain considered here represents a simplified model, whereas realistic devices likely host networks of coupled TLS structures. 
Future work may therefore extend these simulations to larger-scale TLS networks and explore their collective thermalization dynamics.

\section*{Acknowledgements}
We thank the anonymous referees for their careful reading and constructive comments, which helped improve the presentation and clarity of this work.
The numerical simulations in this work made use of the QuSpin package~\cite{Weinberg2017,Weinberg2019}.

\section*{Data Availability}
The numerical simulation data and simulation code supporting the findings of this study are available in the Zenodo database at ~\cite{Guo2026Dataset, Guo2026Software}.

\appendix

\renewcommand{\thefigure}{A\arabic{figure}}
\setcounter{figure}{0}

\section{Additional Results}

\begin{figure*}[htbp]
    \centering
    \includegraphics[width=\linewidth]{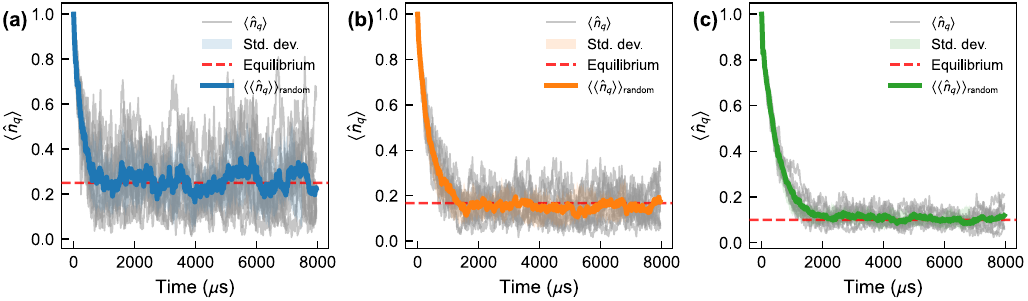}
\caption{
\textbf{Exponential qubit relaxation for different TLS chain lengths.}
All parameters are identical to those used in Fig.~4(c), except for the length of the TLS chain. The initial state of the system is
$|1\rangle_q|0\rangle_{\tau}|1\rangle_{\upsilon}$.
Panels (a), (b), and (c) correspond to TLS chain lengths of 3, 5, and 9, respectively. Consequently, the equilibrium expectation values of the superconducting qubit occupation are $1/4$, $1/6$, and $1/10$, respectively. The energy relaxation times extracted from exponential fits to the data in panels (a), (b), and (c) are 257.4, 381.8, and 519.6~$\mu\mathrm{s}$, respectively.
}\label{figA1}
\end{figure*}

As shown in Fig.~\ref{figA1}, the equilibrium fluctuations become pronounced at very small system sizes. This originates from the restricted accessible Hilbert space, which severely limits the available pathways for coherent transport and quantum evolution. In the extreme limit of a single lattice site, neither wave propagation nor interference can occur and the concept of equilibrium loses its physical meaning. As the accessible state space increases, waves explore increasingly complex propagation pathways, leading to enhanced sensitivity of coherence to perturbations.

Furthermore, we emphasize that random phase perturbations alone cannot drive a single-particle wave into an equilibrium state. Information erasure emerges only after entanglement is established between two (or more) particles. In the framework of quantum three-body chaos, collisions between two particles generate entanglement, whereas subsequent collisions involving the third particle effectively introduce phase perturbations, providing the physical mechanism for information erasure.

\begin{figure*}[htbp]
    \centering
    \includegraphics[width=\linewidth]{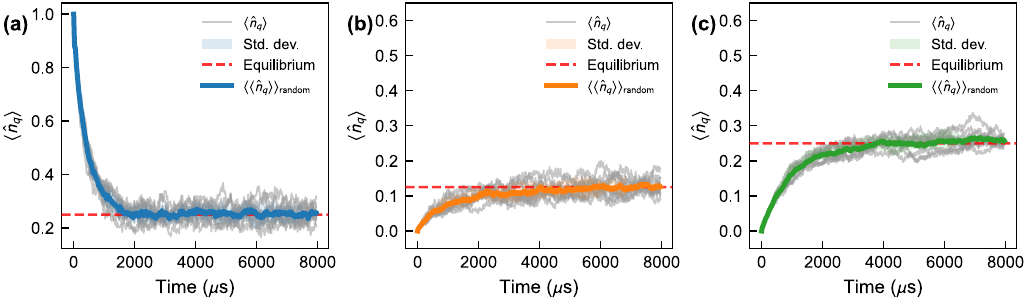}
\caption{
\textbf{Qubit relaxation and thermalization for different TLS excitation numbers.}
All parameters are identical to those used in Fig.~4(c), with a TLS chain length of 7. The initial states of the system in panels (a), (b), and (c) are
$|1\rangle_q|1\rangle_{\tau}|1\rangle_{\upsilon}$,
$|0\rangle_q|1\rangle_{\tau}|2\rangle_{\upsilon}$, and
$|0\rangle_q|2\rangle_{\tau}|2\rangle_{\upsilon}$, respectively.
Consequently, the equilibrium expectation values of the superconducting qubit occupation are $1/4$, $1/8$, and $1/4$, respectively. 
}
\label{figA2}
\end{figure*}

\begin{figure*}[htbp]
    \centering
    \includegraphics[width=\linewidth]{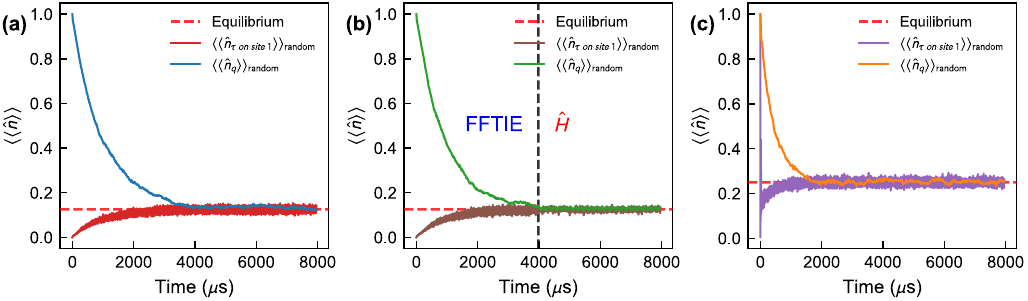}
\caption{
\textbf{Energy relaxation and thermalization from the perspectives of qubit and TLS.}
All parameters are identical to those in Fig.~4(c), with a TLS chain length of 7. The initial states for panels (a), (b), and (c) are $|1\rangle_q|0\rangle_{\tau}|2\rangle_{\upsilon}$, $|1\rangle_q|0\rangle_{\tau}|2\rangle_{\upsilon}$, and $|1\rangle_q|1\rangle_{\tau}|1\rangle_{\upsilon}$, respectively.
(a) The blue curve shows the energy relaxation of the qubit, while the dark red curve denotes the occupation number of the first site of the $\tau$-type TLS chain during thermalization. Both the qubit and the TLS relax toward their equilibrium values.
(b) The green and brown curves correspond to the qubit and the first site of the $\tau$-type TLS chain, respectively. During the first half of the evolution, the system is driven by FFTIE, followed by free evolution under $\hat{H}$. FFTIE drives the system into equilibrium, after which the state remains unchanged under $\hat{H}$. This is because the $\tau$ and $\upsilon$ particles cannot be disentangled by the subsequent evolution under $\hat{H}$ after the information-erasure sequence.
(c) The orange and purple curves correspond to the qubit and the first site of the $\tau$-type TLS chain, respectively. Initially, both are prepared in the excited state. The TLS excitations exhibit rapid coherent oscillations at short times and eventually relax toward equilibrium at long times.
All curves are averaged over ten evolution trajectories, with each individual trajectory exhibiting a similar qualitative behavior.
}
\label{figA3}
\end{figure*}

Figure~\ref{figA2} shows the energy relaxation dynamics of the qubit starting from the $|0\rangle$ and $|1\rangle$ states. It can be seen that $\langle \hat{n}_q \rangle$ always relaxes to its thermal-equilibrium value.
Figure~\ref{figA3} shows the thermalization dynamics of both the qubit and the TLS. Figure~\ref{figA3}(b) shows that the system first reaches thermal equilibrium under the FFTIE sequence and then remains in the equilibrium state during free evolution under $\hat{H}$. This is because the $\tau$ and $\upsilon$ particles become entangled under the FFTIE sequence, and this entanglement cannot be disentangled under $\hat{H}$.

\clearpage
%

\end{document}